\documentclass[a4paper]{aa}
\usepackage{txfonts}
\usepackage[usenames,dvipsnames,svgnames,table]{xcolor}
\usepackage{graphicx}
\usepackage{url}

\usepackage{hyperref}
\usepackage[hyphenbreaks]{breakurl}

\outer\def\gtae {$\buildrel {\lower3pt\hbox{$>$}} \over 
{\lower2pt\hbox{$\sim$}} $}
\outer\def\ltae {$\buildrel {\lower3pt\hbox{$<$}} \over 
{\lower2pt\hbox{$\sim$}} $}
\newcommand{\Msun} {$M_{\odot}$}

\newcommand{\swift}{\sl Swift}
\newcommand{\rosat}{\sl ROSAT}
\newcommand{\suz}{\sl Suzaku}
\newcommand{\xmm}{\sl XMM-Newton}
\newcommand{\asca}{\sl ASCA}

\begin{document}

\title{Distances of CVs and related objects derived from {\sl Gaia} Data Release 1}
\author{Gavin Ramsay\inst{1}, Matthias R. Schreiber\inst{2}, Boris
  T. G\"{a}nsicke\inst{3} \and Peter J. Wheatley\inst{3}}
\authorrunning{Ramsay et al.}
\titlerunning{Distances of CVs derived from {\sl Gaia} DR1}
\institute{Armagh Observatory and Planetarium, College Hill, Armagh, BT61 9DG,
  UK\label{inst1} \and
Instituto de F\'{i}sica y Astronom\'{i}a, Universidad de
 Valpara\'{i}so, Valpara\'{i}so, Chile\label{inst2}\and
Department of Physics, University of Warwick, Coventry CV4 7AL, UK\label{int3}\\
 \email{gar@arm.ac.uk}}

\date{Accepted: April 3 2017}

\abstract {We consider the parallaxes of sixteen cataclysmic variables
  and related objects that are included in the TGAS catalogue, which
  is part of the \textit{Gaia} first data release, and compare these
  with previous parallax measurements.  The parallax of the dwarf nova
  SS\,Cyg is consistent with the parallax determination made using the
  VLBI, but with only one of the analyses of the \textit{HST} Fine
  Guidance Sensor (FGS) observations of this system. In contrast, the
  \textit{Gaia} parallaxes of V603\,Aql and RR\,Pic are broadly
  consistent, but less precise than the \textit{HST}/FGS
  measurements. The \textit{Gaia} parallaxes of IX\,Vel, V3885\,Sgr,
  and AE\,Aqr are consistent with, but much more accurate than the
  \textit{Hipparcos} measurements.  We take the derived \textit{Gaia}
  distances and find that absolute magnitudes of outbursting systems
  show a weak correlation with orbital period. For systems with
  measured X-ray fluxes we find that the X-ray luminosity is a clear
  indicator of whether the accretion disc is in the hot and ionised or
  cool and neutral state. We also find evidence for the X-ray emission
  of both low and high state discs correlating with orbital period,
  and hence the long-term average accretion rate. The inferred mass
  accretion rates for the nova-like variables and dwarf novae are
  compared with the critical mass accretion rate predicted by the Disk
  Instability Model. While we find agreement to be good for most
  systems there appears to be some uncertainty in the system
  parameters of SS\,Cyg. Our results illustrate how future Gaia data
  releases will be an extremely valuable resource in mapping the
  evolution of cataclysmic variables.}


\maketitle

\section{Introduction}

Cataclysmic Variables (CVs) are binary stars containing a white dwarf
that is accreting material from a red dwarf secondary (see Warner 1995
for a comprehensive review).  Their observed characteristics are
inhomogeneous and are largely set by the orbital period, the masses of
the two stars, and the magnetic field strength of the white
dwarf. Angular momentum is gradually lost from the system through
magnetic braking and gravitational wave emission resulting in a
shrinking orbital separation between the stellar components, reaching
a minimum orbital period, $P_\mathrm{orb}\simeq80$\,min (see
G\"ansicke et al. 2009; Knigge, Baraffe \& Patterson 2011).

CVs fall into two broad categories: in those with weakly or
non-magnetic white dwarfs the mass transfer proceeds through an
accretion disc, whereas strong magnetic white dwarfs disrupt the disc
(the DQ Her stars or Intermediate Polars), or prevent its
formation altogether (the AM Her stars or Polars). Among the
non-magnetic systems, the long-term variability further divides
CV subtypes based on the mass loss rate of the companion
star. Nova-like (NL) CVs have mostly $P_\mathrm{orb}>3$\,hr, high mass
transfer rates and stable discs which are hotter than the ionisation
threshold of hydrogen. The NL systems include various
  sub-classes, including the UX UMa systems with steady bright discs,
  and the VY\,Scl systems which undergo low states when the brightness
  drops by 2--5 magnitudes. These episodes are thought to be due to a
  reduction (or cessation) in the mass transfer rate. The dwarf novae
(DN) spend most of the time in a faint quiescent state, in which their
discs are cool and neutral. Occasionally, every few weeks to
  years, they show outbursts, during which the disc is heated above
the ionisation temperature of hydrogen and the accretion rate through
the disc increases dramatically.  A subset of DN (SU\,UMa systems,
generally with $P_\mathrm{orb}\la2$ hr) additionally show longer and
brighter superoutbursts that typically occur every few months.
  Other DN subsets (U Gem and Z Cam systems) typically have
  $P_\mathrm{orb}\ga3$ hr and also show outbursts. In addition the
  Z\,Cam systems show `standstills' in their light curve and are
  thought to be on the boundary between NL variables with their hot
  stable discs, and DN with their unstable discs.

\begin{table*}
\centering
  \caption{16 CVs and related objects from the catalogue of Ritter \&
    Kolb have a parallax measurement in the {\sl Gaia} DR1 TGAS
    catalogue. We arrange them according to the CV sub-type, and
    include their orbital periods, parallax derived from DR1
    (milli-arcsec) and previous parallax information where
    available. References: [1] Harrison et al. (2013); [2] Duerbeck
    (1999); [3] Thorstensen (2003); [4] Harrison et al. (1999); [5]
    Miller-Jones et al. (2013); [6] Nelan \& Bond (2013); [7] Harrison
    \& McArthur (2016); [8] van Altena et al (1995).}
\begin{tabular}{lrcrr}
\hline
Source & P$_{orb}$ & Sub- & Gaia Parallax & Pre-Gaia \\
       &  (hrs)    & Type & (mas)    & Parallax (mas)\\
\hline
\multicolumn{5}{l}{Classical Novae} \\ 
V603 Aql  & 3.32 &N Aql 1918 & 2.92$\pm$0.54 & 4.01$\pm$0.14 [1]\\ 
RR Pic    & 3.48 & N Pic 1925 & 2.45$\pm$0.44 & 1.92$\pm$0.18 [1]\\
HR Del    & 5.14 &N Del 1967 & 0.99$\pm$0.61 & \\
\hline
\multicolumn{5}{l}{Nova-like (NL)} \\ 
TT Ari    & 3.30 &VY Scl & 4.37$\pm$0.42& \\
IX Vel    & 4.65 &UX UMa & 11.28$\pm$0.26 & 10.4$\pm$1.0 [2]\\
UX UMa    & 4.72 &UX UMa  & 3.78$\pm$0.28 & \\
V3885 Sgr & 4.97 &UX UMa  & 7.38$\pm$0.32 & 9.1$\pm$2.0 [2]\\
\hline
\multicolumn{5}{l}{Dwarf Novae (DN)}\\ 
RX And    & 5.04 &Z Cam & 5.41$\pm$0.55 & \\
HL CMa    & 5.20 &Z Cam & 3.19$\pm$0.28 &  \\
AH Her    & 6.19 & Z Cam & 3.02$\pm$0.28 & 3.0$\pm$1.5 [3] \\
SS Cyg    & 6.60 &U Gem  & 8.56$\pm$0.33 & 6.02$\pm$0.46 [4] \\
          &      &       &               & 8.80$\pm$0.12 [5] \\
          &      &       &               & 8.30$\pm$0.41 [6]  \\
          &      &       &               & 7.30$\pm$0.20 [7] \\
Z Cam     & 6.96 &Z Cam & 4.56$\pm$0.24  & 8.9$\pm$1.7 [3] \\
BV Cen    & 14.64 & U Gem & 2.81$\pm$0.38 &  \\  
\hline 
\multicolumn{5}{l}{Magnetic} \\ 
AE Aqr    & 9.88 &DQ Her & 10.95$\pm$0.26 & 9.8$\pm$2.8 [2] \\
\hline
\multicolumn{5}{l}{Others} \\ 
QU Car    & 10.90 & Binary & 2.28$\pm$0.35 & \\
V Sge     & 12.3 &  Binary & -0.35$\pm$0.60 & 3.1$\pm$13 [8]\\
\hline
\end{tabular}
\label{parallax}
\end{table*}

It is now widely accepted that outbursts occur due to instabilities in
the accretion disc, with some role for irradiation of the secondary
influencing the mass transfer rate. This Disc Instability Model (DIM)
and can explain many (but not all) of the features seen in DN and
other outbursting systems (see e.g. H\={o}shi 1979, Meyer \&
Meyer-Hofmeister 1981, and also Lasota 2001 for an extensive
review). A critical observational key test of the DIM is whether it
can correctly predict the recurrence time and the observed
peak and quiescent optical magnitudes of outbursting systems at known
distances.

SS\,Cyg is one of the brightest and best-studied DN in the sky,
reaching $V\simeq8$ at peak outburst, and with a photometric record
extending more than a century (e.g. Cannizzo 2012) and has therefore
been used as a test case for the DIM. Bailey (1981) used the $K$ band
magnitude (see \S 2) to determine a distance to SS Cyg of
95$^{+18}_{-39}$ pc.  Using the {\sl HST} fine guidance sensor to
measure its parallax, Harrison et al. (1999) determined a distance of
166$\pm$12\,pc.  Schreiber \& G\"{a}nsicke (2002) found that for this
distance, the accretion disc limit cycle model (e.g. Cannizzo 1993)
would imply a mean accretion rate too high to undergo outbursts. In
other words, for a distance of 166\,pc, SS\,Cyg should be a NL
variable (see also Schreiber \& Lasota 2007). The distance of SS\,Cyg
has since then been subject to intense debate. Miller-Jones et
al. (2013) obtained a VLBI parallax, which implied a distance of
$114\pm2$\,pc, consistent with the observed outburst properties in the
DIM framework. The original {\sl HST} data has been re-analysed
several times, giving contradictory results (e.g.\ Nelan \& Bond 2013;
Harrison \& McArthur 2016). However, only one of these results is
  consistent with the VLBI measurement (see Table 1).

The first Gaia release (DR1) on 2016 Sept 14 included astrometric and
photometric data of more than one billion stars, and parallax
information for more than two million sources as a result of a joint
{\sl Tycho-Gaia} astrometric solution (TGAS, Brown et al. 2016). We
cross-matched the TGAS catalogue with the Ritter \& Kolb catalogue of
CVs and related objects (Ritter \& Kolb 2003, version 7.23, December
2015) and find 16 objects which are common to both (these are
identified in Table\,\ref{parallax}). They have a range of orbital
period between 3.3--14.6\,hrs and a diversity of sub-type, with three
classical novae which erupted in the early-mid 20th century
(V603\,Aql, RR\,Pic, HR\,Del); four NL variables (IX\,Vel, UX\,UMa,
V3885\,Sgr, and the VY\,Scl system TT\,Ari); six DN (SS\,Cyg, BV\,Cen,
and the four Z\,Cam systems RX\,And, HL\,CMa, AH\,Her, and Z\,Cam);
one magnetic CV (AE\,Aqr) and two systems that were claimed to be CVs,
but are more likely massive binaries which may be interacting
(QU\,Car, V\,Sge).

Since the current data release includes only bright ($V\la13$) stars,
the sample is biased towards CVs with high accretion rates and long
orbital periods.  In this paper we outline their parallaxes and
distances as determined from the {\sl Gaia} DR1, and compare these
with previous measurements. We also derive the absolute magnitudes
of these systems making use of the AAVSO light curves and the TGAS
distances, as well as their X-ray luminosities using fluxes from the
literature or from unpublished {\sl Swift} observations. Finally, we
assess whether the mean mass accretion rates implied by the {\sl Gaia}
distances are consistent with the critical mass accretion rate as
predicted by the DIM.

\begin{table*}
\centering
\caption{We show the derived distances for 16 CVs and related objects
  in Table 1 taken from the catalogue of Astraatmadja \& Bailer-Jones
  (2016) and assumed the scale length, $L=0.11$\,kpc. We also give the
  distance estimates prior to {\sl Gaia} where available, the observed
  optical brightness range in the $V$-band determined from AAVSO data
  spanning the period 1995~--~2015 (apart from the three old novae),
  and the range in absolute magnitude $M_{V}$ implied by the TGAS
  distances. We determine X-ray luminosities using the TGAS distance
  and X-ray flux compiled from the literature or from an analysis of
  {\sl Swift} data.  References: [1] Harrison et al. (2013); [2] Gill
  \& O'Brien (1998); [3] Harman \& O'Brien (2003); [4] G\"{a}nsicke et
  al. (1999); [5] Duerbeck (1999); [6] Baptista et al. (1995); [7]
  Sion et al. (2001); [8] Thorstensen (2003); [9] Miller-Jones et
  al. (2013); [10] Nelan \& Bond (2013); [11] Harrison et al. (1999);
  [12] Sion et al. (2007); [13] Gilliland \& Philips (1982); [14] van
  Altena, Lee \& Hoffleit (1995).}
\resizebox{\textwidth}{!}{
\begin{tabular}{lrcrrrr}
\hline
Source &  Pre-Gaia Distance &  Gaia Distance & AAVSO Vis & $M_{V}$ & X-ray luminosity\\
       &  (pc)              &  (pc) & (mag) & (mag) & (erg s$^{-1}$) \\
\hline
\multicolumn{6}{l}{Classical Novae} \\ 
V603 Aql  &  249$^{+9}_{-8}$ HST parallax [1]& 328.3$\pm$77.9 & --0.5 $\rightarrow$ 11.7 & -8.1 $\rightarrow$ 4.1 & 2.9$\times10^{32}$ ({\swift})\\ 
RR Pic    &  521$^{+54}_{-45}$ HST [1], 600$\pm$60 expansion [2] &  388.3$\pm$87.8 & 1.0 $\rightarrow$ 12.2 & -6.9 $\rightarrow$ 4.3 & 0.4$\times10^{31}$ (0.3--5keV, {\sl XMM})\\
HR Del    &  970$\pm$70 expansion [3] & 560.7$\pm$170.5 & 3.6 $\rightarrow$ 12.1 & -5.1 $\rightarrow$ 3.4 & 1.5$\times10^{31}$ ({\swift})  \\
\hline
\multicolumn{6}{l}{Nova-like} \\ 
TT Ari    &  335$\pm$50 white dwarf/secondary [4] & 228.8$\pm$29.6 & 10.1 $\rightarrow$ 15.9 & 3.3 $\rightarrow$ 9.1 & 6.4--131$\times10^{30}$ ({\swift})\\
IX Vel    &  96$^{+10}_{-8}$ Hipp parallax [5] & 88.8$\pm$3.2 & 9.0 $\rightarrow$ 10.5 & 4.3 $\rightarrow$ 5.8 & 7.9$\times10^{30}$ ({\asca}) \\
UX UMa    &  345$\pm$34 colours/models [6] & 263.5$\pm$30.4 & 12.2 $\rightarrow$ 14.9 & 5.1 $\rightarrow$ 6.7 & 0.7$\times10^{31}$ (0.2--10keV, {\sl XMM})\\
V3885 Sgr &  110$^{+30}_{-20}$ Hipp parallax [5] & 135.9$\pm$8.3 & 9.8 $\rightarrow$ 10.8 & 4.1 $\rightarrow$ 5.1 & 7.7$\times10^{30}$ ({\swift}) \\
\hline
\multicolumn{6}{l}{Dwarf Novae} \\ 
RX And    &  $\sim$200 spectra/models [7] & 185.5$\pm$23.7 & 10.1 $\rightarrow$ 15.6 & 3.8 $\rightarrow$ 9.3 & 3.6--53$\times10^{30}$ ({\swift})\\
HL CMa    & None & 309.8$\pm$42.7 & 10.6 $\rightarrow$ 14.9 & 3.1 $\rightarrow$ 7.4 & 2.5--4.0$\times10^{32}$ ({\swift}) \\
AH Her    &  660$^{+270}_{-200}$ parallax [8] & 325.0$\pm$47.2  & 10.9 $\rightarrow$ 14.8 & 3.3 $\rightarrow$ 7.2 & \\
SS Cyg    &  114$\pm$2 VLA [9]  & 117.1$\pm$6.2 & 8.0 $\rightarrow$ 12.7 & 2.6 $\rightarrow$ 7.4 & 4.7--63$\times10^{31}$ ({\swift})\\
          &  121$\pm$6, 166$\pm$12 HST [10,11] & &  &  & \\
Z Cam     &  163$^{+68}_{-38}$ parallax [8] & 219.3$\pm$19.5 & 10.1 $\rightarrow$ 14.0 & 3.4 $\rightarrow$ 7.3 & 8.1--18.4$\times10^{30}$ ({\swift})\\
BV Cen    & $\sim$435 UV models [12] & 344.3$\pm$64.8 & 10.7 $\rightarrow$ 13.5 & 3.0 $\rightarrow$ 5.8 & 1.9$\times10^{32}$ (2--10keV, {\suz}) \\  
\hline 
\multicolumn{6}{l}{Magnetic} \\ 
AE Aqr    &  102$^{+42}_{-23}$ parallax [8] & 91.4$\pm$3.3 & 10.6 $\rightarrow$ 12.2 & 5.8 $\rightarrow$ 7.4 & 7.5$\times10^{30}$ ({\sl XMM} {\suz}) \\
\hline
\multicolumn{6}{l}{Others} \\ 
QU Car    &  $>$500 [13] & 410.2$\pm$86.1 & 10.8 $\rightarrow$ 12.4 & 2.7 $\rightarrow$ 4.3 & 1.9$\times10^{32}$ (0.2--12 keV {\sl XMM})\\
V Sge     & 320$_{-260}^{+\infty}$ parallax [14] & 761.5$\pm$213.4 & 9.6 $\rightarrow$ 13.3 & 0.2 $\rightarrow$ 2.3 & \\
\hline
\end{tabular}}
\label{distances}
\end{table*}

\section{Determining distances to CVs and related objects}

Since CVs show composite spectra, determining their distances has long
been subject to substantial uncertainties. Early measurements relied
on parallax determinations and proper motion studies for nearby bright
CVs (e.g. Kraft \& Luyten 1965). Only six CVs were included in the
{\sl Hipparcos} catalogue (Duerbeck 1999), and an additional small
number of relatively accurate parallaxes were obtained using
ground-based measurements (Thorstensen 2003, 2008).

Based on the fact that the overall contribution of the white dwarf and
accretion disc to the overall brightness of the CV at infrared
wavelengths is small compared to the secondary star, Bailey (1981)
developed a method using the $K$ band magnitude which was insensitive
to the evolutionary state and temperature of the secondary star. This
method was further refined by Beuermann (2006) who made use of the
surface brightness of the secondary star in selected TiO absorption
bands. An independent estimate of the distance can be obtained for
those CVs where the white dwarf dominates at ultraviolet wavelengths
(e.g. G\"ansicke et al. 2005), however, those estimates are limited by
the unknown white dwarf mass, and hence radius (see
Sect.\,\ref{s-temperature}).

Less precise estimates of CV distances are based on the expansion rate
of the shell in classical novae (e.g. Cohen 1988) and on an empirical
relationship between $M_{V}$ at outburst and orbital period (Warner
1987, see also Patterson 2011). Distances to CVs have also been
determined by comparing the linear polarisation of the CV with the
distance-polarisation relationship (Barrett 1996) for nearby field
stars (where the distances of the field stars are determined from
photometry and spectral type).

\section{Distances determined using {\sl Gaia}}

ESA's {\sl Gaia} mission was launched in 2013 Dec with the goal to map
the sky down to $g\sim20.7$ measuring fundamental parameters such as
position, photometry, parallax and proper motion of at least one
billion stars (Prusti et al. 2016). The {\sl Gaia} first data release
DR1 uses data taken between 2014 July 25 and 2015 Sept 16 (Brown et
al. 2016, Lindegren et al. 2016), and provides $G$-band magnitudes and
positions for 1.1 billion sources. For two million bright stars
included in the Tycho-2 catalogue, parallaxes and proper motions are
calculated using a joint {\sl Tycho-Gaia} Astrometric Solution (TGAS).
We list the {\sl Gaia} parallaxes and uncertainties for 16 CVs and
related objects which are in the TGAS catalogue in Table
\ref{parallax} together with previously determined parallaxes. We
provide a short background summary of each source in the Appendix.

In the upper panel of Fig \ref{comparedistance} we illustrate the
parallaxes of sources shown in Table 1. For the reasons outlined in
\S\,1, the most keenly anticipated parallax of a CV was that of
SS\,Cyg. We find that the {\it Gaia} parallax (8.56$\pm$0.33 mas) is
consistent with that determined using the VLA (8.80$\pm$0.12 mas,
Miller-Jones et al. 2013). However, it is consistent with only one of
the HST parallax measurements (8.30$\pm$0.41, Nelan \& Bond
2013). There are HST parallaxes of two classical novae, V603 Aql and
RR Pic, which are formally consistent with the {\it Gaia}
parallexes at the 2$\sigma$ and 1$\sigma$ level respectively. The
Hipparchos measurements of the NL systems IX Vel and V3885 Sgr
are relatively high, but consistent with those of {\it Gaia} as are
the ground based parallaxes of Thorstensen (2003).

For reasons outlined in Bailer-Jones (2015), obtaining distances and
realistic uncertainties from parallaxes is not a trivial task
(i.e. simply inverting the parallax from TGAS is not appropriate for
systems with parallax uncertainties $\ga20$\,\%). Here, we make use of
the distances and errors from the catalogue published by Astraatmadja
\& Bailer-Jones (2016), which require a choice of the scale length
$L$. We compared the distances assuming $L=0.11$\,kpc and 1.35\,kpc,
and find that for sources closer than $\sim$500pc the distances and
uncertainties are very similar. However, for more distant sources the
distances and errors determined assuming $L=1.35$\,kpc start to become
significantly larger compared to those determined assuming
$L=0.11$\,kpc. In the following, we adopt the distance based on the
choice of $L=0.11$\,kpc (which Astraatmadja \& Bailer-Jones 2016
determined by fitting their Bayesian prior with the true distance
distribution of stars included in a model of bright stars in the Milky
Way). We list the derived {\sl Gaia} distances and the pre-{\sl Gaia}
distances in Table \ref{distances}, where we indicate whether the
source's pre-{\sl Gaia} distance was determined using a parallax
measurement or through modelling of spectra or colours. We illustrate
this comparison in Figure \ref{comparedistance}.

We now make a more detailed comparison of the distances of the CVs and
related objects in our sample with previous distance measurements. The
distance and uncertainty to SS Cyg is virtually identical for the
three scale lengths tabulated by Astraatmadja \& Bailer-Jones (2016):
$117.1\pm6.2$\,pc. This result is at odds with that of the most recent
analysis of the {\sl HST} derived parallax ($137\pm4$\,pc; Harrison \&
McArthur 2016). However, it is in excellent agreement with the
parallax measurement using the VLBI ($114\pm2$\,pc; Miller-Jones et
al.\ 2013) whose results were held up as a demonstration of the DIM
being able to reproduce the optical outburst characteristics of this
DN (see for example Schreiber \& G\"{a}nsicke 2002). However, the
apparent discrepancy of the various distance measurements meant that
the question of whether the DIM was able to reproduce the
characteristics of the long term light curve of SS\,Cyg (and hence of
DN more generally) remained controversial. The {\sl HST} study which
is closest to (and consistent with) the {\sl Gaia} distance is that of
Nelan \& Bond (2013) who found a distance of $120.5\pm5.7$ pc.  We add
in passing that Schreiber \& G\"{a}nsicke (2002) note at the end of
their paper that a distance of $\sim$117 pc would make the mass
transfer rate derived from observations agree with the DIM. The
  {\sl Gaia} distance to SS Cyg is 117.1$\pm$6.2\,pc and matches precisely with
  the prediction of the DIM. We show in Table \ref{parallax} the
parallax measurements derived by the groups discussed above.

\begin{figure}
\begin{center}
\setlength{\unitlength}{1cm}
\begin{picture}(16,12.)
\put(-0.6,0.){\includegraphics{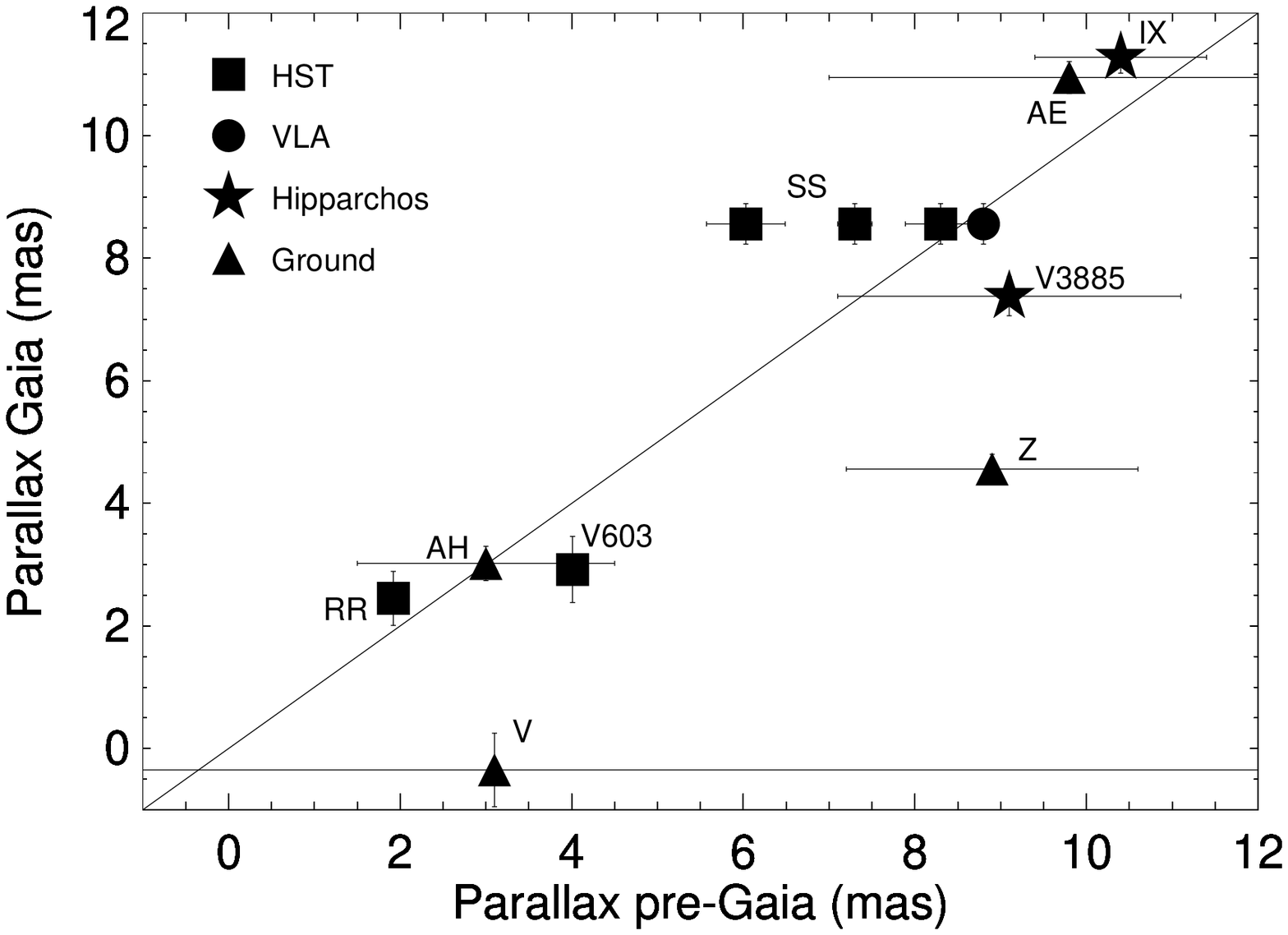}}
\put(-0.6,-6.3){\includegraphics{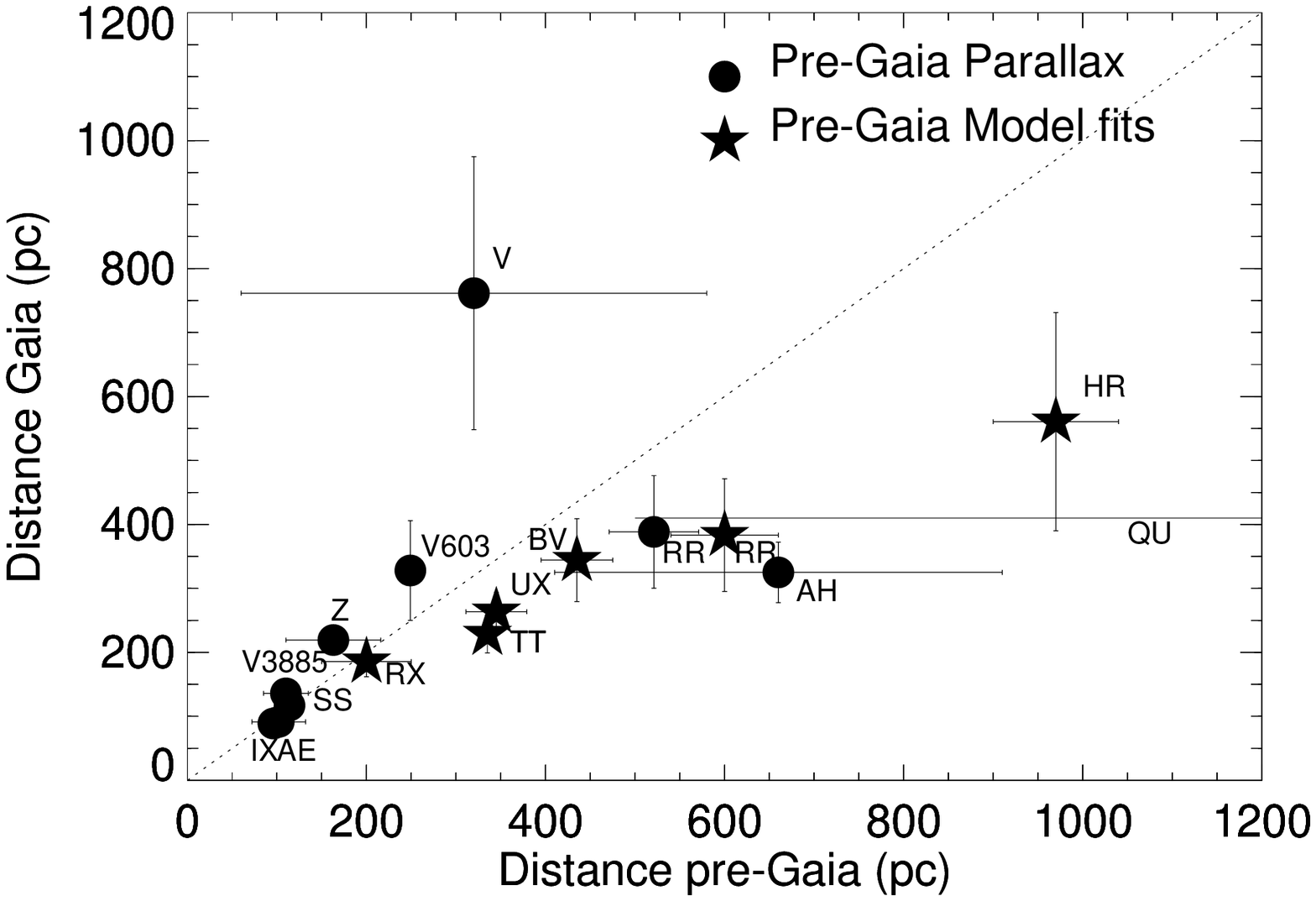}}
\end{picture}
\end{center}
\caption{In the upper panel we illustrate the parallaxes of sources
  which are in the TGAS catalogue and had previously determined
  parallaxes (the data measurements are outlined in Table 1). For SS
  Cyg there are four previous parallaxes. In the lower panel we
  compare the distances of the 16 CVs in TGAS (taken from Astraatmadja
  \& Bailer-Jones 2016 and tabulated in Table 2) with the best
  distance estimate pre-{\sl Gaia}. We indicate by different symbols
  whether the pre-{\sl Gaia} distance was determined via a parallax
  measurement or modelling spectra or colours. We note which system
  corresponds to which point by means of an abbreviation of the source
  name.}
\label{comparedistance} 
\end{figure}

Two other systems, the old novae V603\,Aql, and RR\,Pic have {\sl HST}
parallaxes which imply a distance of $249^{+9}_{-8}$\,pc and
$521^{+54}_{-45}$\,pc for V603 Aql and RR Pic respectively. As
indicated previously, the errors on the {\sl Gaia} distances are quite
large and the distances derived using {\sl HST} and {\sl Gaia} are
formally consistent. CVs with distances determined by {\sl Hipparcos}
(IX\,Vel, V3885\,Sgr and AE\,Aqr) are consistent with the distances
determined using {\sl Gaia}. The very uncertain ground-based parallax
of AH\,Her is consistent with, but now superseded by the lower, and
much more accurate {\sl Gaia} distance. For Z\,Cam, the best
ground-based parallax is replaced by a somewhat larger {\sl Gaia}
distance. QU\,Car has a {\sl Gaia} distance which is consistent with
the lower limit previously found but not with the larger distance
estimate of 2\,kpc (Drew et al 2003) and is therefore not as luminous
as was once thought. With a distance of 760\,pc the massive binary
V\,Sge is the most luminous source under study here (with the
exception of the old novae during their eruptions).

The other two old novae, RR Pic and HR Del, have distances determined
using assumptions based on the expansion velocity, which are biased
towards slightly greater distances compared to {\sl Gaia}
distances. For those objects whose distances were determined by
fitting spectra or photometric colours, there is a slight bias for the
{\sl Gaia} distances being closer than the pre-{\sl Gaia}
estimates. In the Appendix we note the distances determined by
Barrett (1996) for individual sources. In summary the distances to
three NL systems are suprisingly consistent with the {\sl Gaia}
distances, although the distances to the classical novae are
underestimated, and the distance to V Sge is grossly underestimated.

\section{Absolute magnitude of outbursting CVs}

We extracted the visual light curves of the 16 CVs in TGAS from the
AAVSO\footnote{\url{http://aavso.org}} database (Kafka 2015) over the
time interval 1995--2015, thereby ensuring the range of their
variability is well characterised (we employed a light clipping to
remove outlying points). The visual measurements generally well sample
this 20 year interval and we found fairly good agreement between
visual measurements, and $V$-band CCD measurements which amateur
astronomers were starting to use more commonly in the second half of
this time period. We list in Table\,1 the range in the visual
magnitude for each source and the absolute magnitude computed using
the TGAS distance (we do not account for interstellar extinction, but
if $E_{B-V}=0.03$, then $M_{V}$ would be brighter by $\sim$0.1 mag).

In Figure\,2 we show $M_{V}$ (max,min) for our targets as a function
of orbital period. We also show the linear fits to outbursting sources
as derived by Warner (1987) using those CVs which had known distances
(Patterson 2011 shows a fit which is very similar to the Warner
relationship). We find that the NL variables IX\,Vel and V3885 Sgr
have $M_{V,\mathrm{max}}$ which are within $\simeq$0.2 mag of the
linear fit of Warner (1987). In contrast, TT\,Ari is $\simeq$1.5 mag
brighter and UX\,UMa is $\simeq$0.7 mag fainter than this relationship
implies. This may be due to an inclination effect since TT Ari has a
low inclination ($\sim30^\circ$) making it appear brighter, whilst
UX\,UMa has a high inclination ($\simeq70^\circ$), making it fainter
(see Patterson 2011 for a discussion on the relationship between
$M_{V}$ and binary inclination). The DN in our sample show no clear
dependence on orbital period with a $M_{V,\mathrm{max}}\sim3.2\pm$0.4,
but we have a smaller sample, and a more restricted range in orbital
period than the samples of Warner (1987) and Patterson (2011). The DN
in our sample have a mean $M_{V,\mathrm{max}}$ which is
$\simeq0.8\pm0.4$ mag brighter than predicted by the relationship of
Warner (1987).

We also show the linear fits to $M_{V,\mathrm{min}}$ derived by Warner
(1987) for outbursting systems in Figure 2. As might be expected the
NL systems (which do not show regular outbursts like the DN) show on
average a $M_{V,\mathrm{min}}$ which is brighter by 1.5$\pm$1.5 mag than
expected from the Warner (1987) relationship. TT\,Ari shows dramatic
dips in its long term light curve (hence the VY\,Scl sub-type). In
contrast, DN are within 0.0$\pm$0.7 mag of the $M_{V,\mathrm{min}}$
orbital period relationship. The one system which is discrepant is RX
And which is 1.3 mag fainter than expected. RX And is an interesting
system in that it appears to show evidence for both Z Cam and VY Scl
behaviour making it fainter when undergoing a dip event (Schreiber,
G\"{a}nsicke \& Mattei 2002).

We note that $M_{V,\mathrm{min}}$ for QU\,Car is consistent with
Warner's linear relationship for outbursting CVs, while for V\,Sge
$M_{V,\mathrm{min}}$ is $\sim$2 mag brighter than the CV trend. This
indicates that V Sge is unlikely to be a CV, whilst the situation for
QU Car is less clear. The magnetic CV AE\,Aqr has an
$M_{V,\mathrm{min}}$ which is $\sim$1 mag fainter than the Warner
(1987) fit for $M_{V,\mathrm{min}}$. This is not unexpected since it
does not have a regular accretion disk (see the next section).

In Table\,2 we also list the observed peak magnitude (taken from
Strope et al. 2010) of the three old novae included in the TGAS
catalogue and determine the outburst $M_{V}$ (also given in Table\,1)
using the {\sl Gaia} distances.  We compare these absolute magnitudes
with the predicted values using the nova calibration of Della Valle \&
Livio (1995), and the observables from Strope et al. (2010). The
predicted values are $M_{V}=-8.9$, --7.0 and --6.8 for V603\,Aql,
RR\,Pic and HR\,Del, respectively, which compares with --8.1, --6.9
and --5.1 before accounting for extinction.  Taking $E_{B-V}=0.08$,
0.0 and 0.17 for V603\,Aql, RR\,Pic and HR\,Del, respectively
(Selvelli \& Gilmozzi 2013), we obtain corrected absolute magnitudes
of $M_{V}=-8.4$, --6.9 and --5.6. The observed magnitude of RR Pic is
within 0.1 mag of the predicted magnitude, but V603 Aql and HR Del
were 0.5 and 1.2 mag fainter than the nova decline calibration
predicts. Given that the error on the {\sl Gaia} distances to these
three novae are 20--30 percent, we await more precise parallax
measurements which will be available in DR2 before making definitive
conclusions regarding the reliability of the nova distance
calibrations.

\begin{figure}
\begin{center}
\setlength{\unitlength}{1cm}
\begin{picture}(16,6.)
\put(10,-0.5){\includegraphics{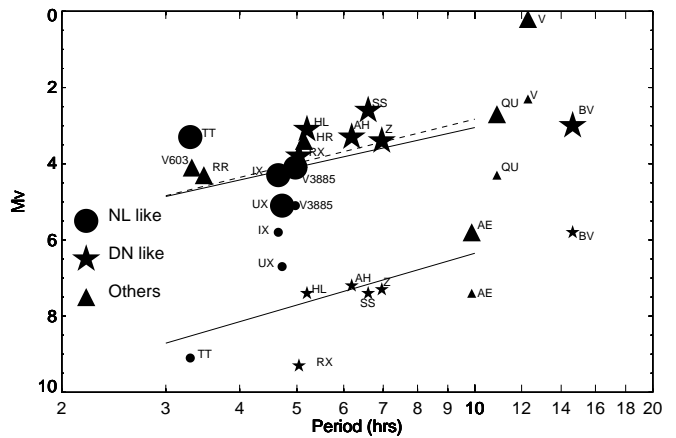}}
\end{picture}
\end{center}
\caption{The absolute magnitude for the maximum and minimum brightness
  for the sources in this study as a function of orbital period. We
  show the relationship of Warner (1987) for $M_{V}$ at maximum and
  minimum where the smaller symbols indicate $M_{V}$ minimum (the
  dashed line indicates the correlation determined by Patterson
  2011). We note which system corresponds to which point by means of
  an abbreviation of the source name.}
\label{Mv} 
\end{figure}

\section{X-ray luminosities}

CVs typically exhibit strong X-ray emission that is seen to
anti-correlate with the optical state of the disc (e.g.\ Mukai
2017). Observations over the outburst cycle of DN, for instance, show
that X-ray emission is brightest during quiescence (e.g. Wheatley et
al. 1996; Collins \& Wheatley 2010, Fertig et al., 2011) and the
X-rays are suppressed during the optical outburst where they are
replaced by intense extreme ultraviolet emission (e.g. Wheatley,
Mauche \& Mattei 2003). This behaviour means that X-ray luminosities
measured with reliable distances can be used to assess the accretion
state of the disc.

The X-rays are known to be emitted from very close to the white dwarf
surface (e.g.\ Mukai et al. 1997; Wheatley \& West 2003) and they are
usually thought to arise in a narrow boundary layer between the
accretion disc and white dwarf surface, where the kinetic energy of
the disc material is thermalised. Indeed, unless the white dwarf is
rotating very rapidly, this boundary layer is expected to emit half of
the total accretion luminosity of the system (Pringle 1981). In order
to explain some of the outburst properties of DN, however, it has also
been suggested that the inner accretion disc may be eroded or
truncated in low accretion rate systems (e.g.\ Livio \& Pringle 1992;
Meyer \& Meyer-Hofmeister 1994; King 1997), in which case the X-ray
emission could account for even more than half of the accretion
luminosity in the low state.

The anti-correlation between X-rays and the optical state of the disc
is usually interpreted as the response of the boundary layer to
changing accretion rate (Pringle \& Savonije 1979). At low accretion
rates the boundary between disc and white dwarf will have low density
and be optically thin to its own emission (whether it is a narrow
boundary layer or a truncated inner disc region). It therefore cools
inefficiently and emits thermal X-rays at high temperatures (Patterson
\& Raymond 1985a). In this state the X-ray luminosity can be used to
trace the total accretion rate onto the white dwarf. At high accretion
rates the density in the boundary region increases and it eventually
becomes optically thick to its own emission, cooling to much lower
temperatures despite its increase in total luminosity (Patterson \&
Raymond 1985b). In this state the luminosity is dominated by
extreme-ultraviolet emission, which is difficult to use as a tracer of
the accretion rate because it is strongly absorbed in the interstellar
medium. However, the X-ray luminosity immediately prior to this
transition can be used to determine the accretion rate at which the
boundary layer becomes optically thick, at least in the case of SS~Cyg
(Wheatley, Mauche \& Mattei 2003).

The accretion rates of quiescent DN inferred from X-ray luminosities
have been problematic for the DIM. This is because the X-ray
luminosities are much higher than is consistent with a low-state disc
extending down to the white dwarf surface (e.g.\ Lasota 2001;
Wheatley, Mauche \& Mattei 2003) and because the accretion rates are
seen to decrease rather than increase between outbursts (McGowan et
al.\ 2004; Collins \& Wheatley 2010; Fertig et al.\ 2011). These
observations tend to support the suggestions that the inner disc is
eroded in the low state.

Previous studies of X-ray luminosities based on parallax distances
have found evidence for a weak correlation with orbital period
(Baskill et al.\ 2005; Byckling et al.\ 2010). This suggests a
correlation also with the long-term average accretion rate that is
driven by the secular evolution of the system. Consistent with this,
there is also a correlation between the quiescent X-ray luminosity and
the outburst duty cycle, where duty cycle is defined as the proportion
of time spent in outburst (Britt et al.\ 2015).

\begin{figure}
\begin{center}
\setlength{\unitlength}{1cm}
\begin{picture}(16,6)
\put(10,-0.5){\includegraphics{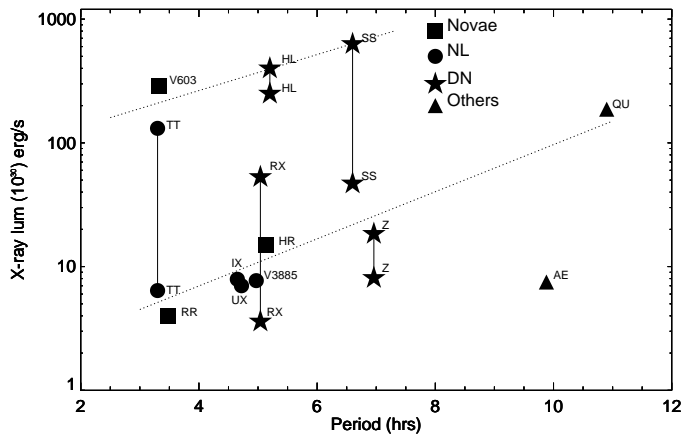}}
\end{picture}
\end{center}
\caption{The X-ray luminosity of the 16 CVs in TGAS as a function of
  orbital period. For some systems we give the range of luminosity as
  quoted in the literature or measured from {\sl Swift} data. We show
  different classes of CV with different symbols. The dotted lines
  illustrate how the X-ray luminosity appears to be related to the
  orbital period in high and low accretion states.}
\label{xraylum} 
\end{figure}

We have identified the X-ray flux of sources from the literature (see
the Appendix for references) and determine their X-ray luminosity
using the TGAS distance (Table 2). For a number of sources we have
extracted the X-ray data from the {\sl Swift} archive and extracted a
mean X-ray spectrum, which we have fitted using an absorbed
two-temperature thermal plasma model (Table 2 indicates which
satellite has been used). In the interests of brevity we do not give
details of the {\sl Swift} observations here, but they have been
reduced and analysed in an identical manner to that outlined for the
study of the symbiotic nova AG Peg (Ramsay et al. 2016). In Table 2 we
give the range in X-ray luminosity where there is more than one epoch
of observation and illustrate this in Figure \ref{xraylum} where we
distinguish between the different classes of CV.

The maximum X-ray luminosities of DN, the old nova V603 Aql, and the
nova-like TT Ari shown in Fig. \ref{xraylum} appear to define an upper
bound to the X-ray luminosity that increases with orbital period from
$L_{X}\sim2\times10^{32}$ erg s$^{-1}$ at 3 h to around
$L_{X}\sim6\times10^{32}$ erg s$^{-1}$ at 7 h. This appears to provide
further support for a correlation between quiescent accretion rate and
the long-term average rate. For TT Ari this maximum X-ray luminosity
corresponds to one of the dips in optical brightness to the quiescent
disc state.

The other old novae (RR Pic and HR Del) and NL (together with TT Ari
in its usual high state) all have lower X-ray luminosities that are in
line with the dwarf novae in outburst. This indicates that the
accretion discs in these novae and NL are in sustained high states,
with high accretion rates onto the white dwarf and optically-thick
boundary layers. It seems that there may also be a dependence of these
high-state X-ray luminosities on orbital period, despite the
considerable scatter seen in the dwarf novae (perhaps reflecting
short-term variability).

AE~Aqr has a low X-ray luminosity, similar to the high accretion rate
states of short period systems, but it is believed to be in a highly
unusual state in which most of the mass transfer flow from the
secondary star is propelled away from the white dwarf by its rapidly
rotating magnetic field (Wynn, King \& Horne 1997). Thus the low X-ray
luminosity in this case is thought to reflect a low accretion rate
onto the white dwarf rather than a high-state disc. The combination of
faint absolute magnitude and low X-ray luminosity seems to be a strong
indicator of the propelling state, and {\sl Gaia} DR2 may allow other
such systems to be identified.

QU~Car has a high X-ray luminosity compared to other systems with
high-state discs, with an X-ray luminosity comparable to quiescent
DN. However, it has a much longer orbital period than the other
systems, and its X-ray luminosity seems to be consistent with an
extrapolation of the possible trend of increasing high-state
luminosity with orbital period (as indicated by the lower dashed line
in Fig.\,3). This is a further indication that the X-ray luminosity of
high accretion rate discs is correlated with the long-term average
mass transfer rate, even though the total boundary layer luminosity is
probably dominated by optically-thick emission in the
extreme-ultraviolet.

Finally, we note that the reduced distance of SS~Cyg has the effect of
reducing the accretion rate at which the boundary later appears to be
become optically-thick to its own emission from $1.0\times10^{16}\,\rm
g\,s^{-1}$ found by Wheatley, Mauche \& Mattei (2003) to
$5.0\times10^{15}\,\rm g\,s^{-1}$, which is closer to that inferred
for other systems by Fertig et al.\ (2011).

\section{Implications for the white dwarf mass determinations}
\label{s-temperature}

The accreting white dwarf dominates the ultraviolet emission of a
significant number of CVs (e.g. Matteo \& Szkody 1984) and model
atmosphere fits to {\sl IUE}, {\sl FUSE}, and {\sl HST} have resulted
in measurements of the white dwarf effective temperatures of
$\simeq75$ CVs (Townsley \& G\"ansicke 2009, Pala et al. 2017), which
provide constraints on the secular mean accretion rates (Townsley \&
Bildsten 2003). However, given that {\sl IUE} and {\sl HST}
observations only cover a relatively small wavelength range, those
measurements are subject to some degeneracy with the surface gravity,
and hence the white dwarf mass. Accurate parallaxes break this
degeneracy, as the flux scaling factor between the data and the model
constrains the white dwarf mass, and via use of a mass-radius relation
(e.g. Panei et al. 2000), the surface gravity. Among the CVs in {\sl
  Gaia} DR1 only TT\,Ari and RX\,And have white-dwarf dominated
ultraviolet spectra.

Two {\sl IUE} spectra were obtained during a prolonged low state of
TT\,Ari in December 1982 and December 1983, which G\"ansicke et
al. (1999) modeled with a hot, $T_\mathrm{eff}\simeq39\,000$\,K, white
dwarf. Adopting a very wide range for the white dwarf mass,
$0.35-1.20\,M_\odot$, they concluded that the distance of TT\,Ari
should be in the range of $125-385$\,pc. Using the {\sl Gaia}
measurement of $\simeq230$\,pc implies that the white dwarf mass is
$\simeq0.9\,M_\odot$, which is close to the mean mass of CV white
dwarfs, $<M_\mathrm{wd}>=0.83\pm0.23\,M_\odot$ (Zorotovic et
al. 2011). Unfortunately, the low quality of the {\sl IUE} spectrum is
now the limiting factor in the accuracy of the white dwarf parameters.

An {\sl HST} spectrum of RX\,And was obtained during an unusual low
state in December 1996, which Sion et al. (2001) modelled with a
34\,000\,K white dwarf. From the flux scaling factor, they estimated a
distance of 200\,pc for a $0.8\,M_\odot$ white dwarf, which is, within
the uncertainties, consistent with the {\sl Gaia} DR1 distance of
$185.5\pm23.7$\,pc. Taking the slightly lower DR1 distance value at
face value would push the white dwarf mass up to $\simeq0.9\,M_\odot$.

These two examples illustrate how accurate distances will help
constrain CV white dwarf masses. The {\sl Gaia} DR2 will not only
provide parallaxes to several dozen CVs which have high-quality
ultraviolet observations, but also reduce the uncertainties in the
parallax measurements, which will substantially increase the number of
CV white dwarfs with accurate ($\le10$\%) mass determinations.

\section{Implications for the disk instability model}

As briefly outlined in \S\,1, the generally accepted model for DN
outbursts is the DIM (Meyer \& Meyer-Hofmeister 1981, Hameury et
al. 1998). According to this model, DN outbursts are produced by a
thermal and viscous instability that does not allow the disk to
transport material inwards at the same rate as it receives it from the
companion star.  Instead, the disk has to switch between an excess of
mass accretion, the so-called high state, and the quiescence state
when matter accumulates in the disk as less mass is accreted than the
disk receives from the secondary.

The DIM explains the observed characteristics of normal DN outbursts
reasonably well. The DIM, however, also faces problems, e.g. it seems
to be difficult to reproduce irregular outbursts (e.g. Schreiber et
al. 2002) and to explain the superoutbursts observed in SU\,UMa
systems the DIM needs to be complemented either by a change in the
mass transfer rate or a tidal instability (Osaki 1989, Smak 2004,
Schreiber et al. 2004). Most of these problems are probably related to
our limited understanding of the quiescence state (see Lasota 2001 for
a review). In quiescence, matter must accumulate in the disk but it is
uncertain at which radius this accumulation is most efficient. Also,
as outlined in \S\,5, quiescent X-ray luminosities are higher than
predicted by the DIM, implying that more material is transported onto
the white dwarf than expected. Together with the observed decrease in
X-ray luminosity between outbursts, this suggests the inner disc is
unexpectedly eroded during quiescence. It seems that the $\alpha$
viscosity prescription used in the DIM does not provide a good
description of quiescence.

In contrast, the $\alpha$ formalism works well in the high state and
the DIM makes clear and testable predictions. As the instability
causing the outbursts is generated by the partial ionisation of
hydrogen, one of the key predictions of the DIM is that above a
certain mass transfer rate, the disk should be fully ionised and
stable and no DN outbursts should occur. This critical mass transfer
rate depends sensitively on the radius of the disk and the mass of the
white dwarf. A frequently used prescription for the critical mass
transfer rate is
\begin{equation}\label{eq-Mcrit}
\dot{M}_{\mathrm{crit}}=9.5\times10^{15} R_{10}^{2.68}
M_{\mathrm{WD}}^{-0.89} \mathrm{g/s}
\end{equation}  
where $M_{\mathrm{WD}}$ is the white dwarf mass and $R_{10}$ the disk
radius in units of $10^{10}$cm (Schreiber \& G\"{a}nsicke 2002). NL
variables and Z\,Cam stars during stand-still, and DN at the end of
long outbursts should reach a quasi stationary state with a mass
accretion rate $\sim\dot{M}_{\mathrm{crit}}$. During these quasi
stationary states, the surface density and effective temperatures in
the disk follow well-defined radial dependencies. In NL variables the
disk must accrete at a rate exceeding the critical value and is
expected to be in a quasi stationary state. At the end of a long DN
outburst or during the stand-stills of Z\,Cam systems, the disk should
be in the quasi stationary state with a mass accretion rate very
similar to the critical rate. For a given set of system parameters, it
is therefore possible to estimate the {\em{absolute}} brightness of a
Z\,Cam disk during stand-still and from DN at the {\sl end of long
  outbursts} as predicted by the DIM. For NL variables the brightness
corresponding to the critical accretion rate represents a lower limit.

These DIM predictions can be confronted with observations if the
system parameters of a given CV and its distance are known reasonably
well. We discussed the case of SS\,Cyg in \S 1, and noted that
Schreiber \& G\"{a}nsicke (2002) found strong disagreement between DIM
predictions and a distance to SS\,Cyg of 166\,pc as measured by
Harrison et al. (1999). For agreement with the DIM, Schrieber \&
G\"{a}nsicke (2002) estimated a distance of $\sim117$\,pc in perfect
agreement with the {\sl Gaia} measurement.

In what follows we confront the DIM prediction with the {\sl Gaia}
distance measurements for 9 CVs, i.e. the four NL variables systems
and 5 of the 6 DN. The long orbital period DN BV\,Cen does not qualify
for the test as its outbursts are relatively short and the system
never reaches the quasi stationary state. For the other 9 systems we
can assume a steady state accretion disk during high-states and
calculate two mass transfer rates. First, we calculate the critical
mass transfer rate as given by Eq.\ref{eq-Mcrit}. This is straight
forward and the main uncertainties involved are the white dwarf and
secondary mass which together determine the disc radius which in the
high-state should be close to the tidal truncation radius. Second, we
calculate the accretion rate that is required to reproduce the
observed absolute brightness assuming a steady state accretion
disk. The uncertainties here are the stellar mass, inclination,
observed brightness, and distance. We take into account all the
uncertainties involved in the determination of both mass accretion
rates. Thanks to the precise {\sl Gaia} measurements, the
uncertainties implied by the distances are minor. The dominating
uncertainties are those of the system parameters which are notoriously
difficult to measure in CVs with high mass transfer rates.

\begin{figure}
\begin{center}
\setlength{\unitlength}{1cm}
\begin{picture}(16,6.5)
\put(0.0,-0.2){\includegraphics{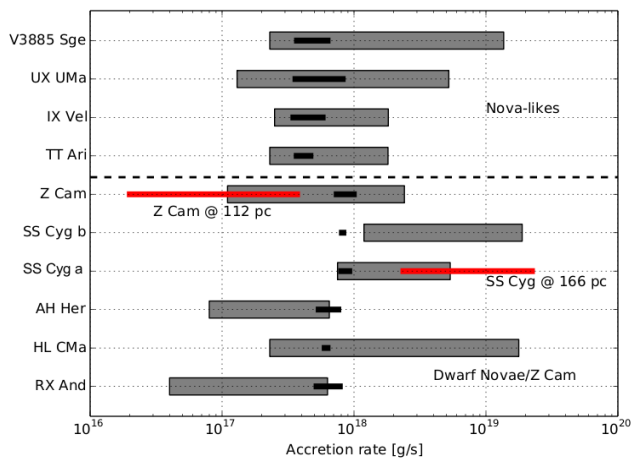}}
\end{picture}
\end{center}
\caption{Comparison of the mass accretion rates derived from
  observations (grey) and the critical mass accretion rate predicted
  by the DIM (black bars). In NL variables the DIM predicts the
  systems to accrete at a rate exceeding the critical one which is
  clearly possible as the grey bars extend to larger accretion rates
  than the black ones. For SS\,Cyg we tested the DIM prediction for
  two sets of parameters [a] and [b] and find agreement only if we use
  those derived by Friend et al. (1990) [a], which allow for a lower
  system inclination and a higher white dwarf mass. For Z\,Cam and
  SS\,Cyg we also show the accretion rates derived from the
  observations using the pre-{\sl Gaia} distances (red bars) to
  demonstrate that the precise measurements now available brought into
  agreement theory and observations. The case of the two Z\,Cam stars
  AH\,Her and RX\,And is slightly worrying for the DIM as in both
  cases the overlap between both accretion rates is marginal. }
\label{fig-DIM} 
\end{figure}

Figure \ref{fig-DIM} illustrates our results. The grey bars represent
the accretion rates derived from observations using the parameters
given in Table\,\ref{Tab-DIM} while the DIM predictions are
represented by the smaller black bars. The grey bars cover much larger
ranges because of the uncertainty of the system parameters. For all NL
variables in our sample, the {\sl Gaia} distances lead to reasonable
agreement with the DIM. The mass accretion rate derived from
observations cover ranges largely exceeding the critical rate as
required by the DIM.  In the case of the Z\,Cam stars and the DN
SS\,Cyg the situation is more complex and we therefore discuss each
system individually.

For Z\,Cam itself the accretion rate we derived using the {\sl Gaia}
distance perfectly agrees with the DIM prediction, i.e. both accretion
rates overlap. Interestingly, the pre-{\sl Gaia} distance of 112\,pc
was in clear disagreement with the model (red bar in
Fig.\,\ref{fig-DIM}). If Z\,Cam was that close, the accretion rate
during stand-still would be below the critical rate and the DIM would
predict DN outbursts instead of a stand-still.

Surprisingly, the only pure DN in our sample, SS\,Cyg, is again
causing some trouble. As mentioned above, the {\sl Gaia} distance is
in excellent agreement with the predictions made by Schreiber \&
G\"{a}nsicke (2002). However, back in 2002 we used the system
parameters derived by Friend et al. (1990) which have been revised
later by Bitner et al. (2007). According to the latter, the white
dwarf in SS\,Cyg is less massive and the inclination
higher than previously thought. Both these changes
require a higher mass accretion rate to reproduce the same
brightness. This higher mass accretion rate is in disagreement with
the DIM as it significantly exceeds the critical value at the onset of
the decline. In Figure\,\ref{fig-DIM} we show the predicted mass
accretion rate and the one derived from observations for both
parameter sets. We also add a red bar which represent the mass
accretion rate if SS\,Cyg was at 166\,pc.

For HL\,CMa the observations agree reasonably well with the DIM
prediction. The mass accretion rate derived from observations covers a
large range of values as the inclination of the system is
unknown. This large range includes the critical mass accretion rate
and thus HL\,CMa might accrete during stand-still at a rate similar to
or slightly larger than the critical one, as predicted by the DIM.

The situation is somewhat worrying for the remaining two systems of
the Z\,Cam type, i.e. AH\,Her and RX\,And. Despite our conservative
uncertainty estimates, the accretion rate derived from observations
hardly reaches the critical mass accretion rate. The system parameters
for both systems are too uncertain to call this a major problem of the
DIM, but if the white dwarf masses can be confirmed to be relatively
large and if the inclination does not greatly exceed the value given
in Table\,\ref{Tab-DIM}, the model might face some serious problems.

\begin{table*}
\centering
\caption{System parameter and brightness values used for testing the
  predictions of the disk instability model. The visual brightness is
  taken from the AAVSO light curve during stand-still of Z\,Cam stars
  and at the end of long outbursts for SS\,Cyg. At these phases the
  DIM predicts steady state accretion at rates close to the critical
  accretion rate. For the NL we used an average brightness. For
  HL\,CMa several parameters are completely unconstrained and we
  therefore used a broad range of inclinations (30--70 degrees) and
  the average WD mass for CVs of 0.83$\pm$0.1. For SS\,Cyg we used two
  different sets of parameter [a] and [b]. If no error for the masses
  is given, we assumed an uncertainty of 0.1$M_{\odot}$. References:
  [1] Linnell et al. (2009); [2] Neustroev et al. (2011); [3] Linnell
  et al. (2007); [4] G\"{a}nsicke et al. (1999); [5] Wu et al. (2002);
  [6] Hartley et al. (2005); [7] Friend et al. (1990); [8] Bitner et
  al. (2007); [9] Horne, Wade \& Szkody (1986); [10] Hutchings et
  al. (1981); [11] Shafter (1983)}
    \label{Tab-DIM}
  \begin{tabular}{lcccrr} 
        \hline
        Name & $M_{\mathrm{WD}}$[$M_{\odot}$] & $M_{2}$[$M_{\odot}$] & inc
        [degree] & $m_{\mathrm{V}}$ & Ref\\
        \hline
        V3885\,Sge & $\sim$0.7 & $\sim$0.475 & $\sim$65 & 9.5$\pm$0.2 &[1]\\
        UX\,UMa & 0.9$\pm$0.3 & 0.39$\pm$0.15 & 70$\pm$5 & 13$\pm$0.2 & [2]\\
        IX Vel & 0.8$\pm$0.2 & 0.52$\pm$0.10 & 57$\pm$2 & 9.5$\pm$0.2 & [3]\\
        TT\,Ari & $\sim$0.9 & $\sim$0.2 & $\sim$30 & 10.8$\pm$0.2 & [4,5]\\
        Z\,Cam &  0.99$\pm$0.15 & 0.71$\pm$0.10 & 57$\pm$11 & 11.5$\pm$0.2 & [6]\\
        SS\,Cyg [a]& 1.19$\pm$0.05 & 0.7$\pm$0.1 & 37--53 & 8.6$\pm$0.2 & [7]\\
        SS\,Cyg [b]& 0.81$\pm$0.2 & 0.55$\pm$0.13 & 45--56 & 8.6$\pm$0.2 & [8]\\
        AH\,Her & 0.95$\pm$0.1 & 0.76$\pm$0.08 & 46$\pm$3 & 12.5$\pm$0.2 & [9]\\
        HL\,CMa & $\sim$0.83$\pm0.1$ & 0.45$\pm$0.1 & $\sim$45 & 11.5$\pm$0.2 & [10] \\
        RX\,And &  1.14 $\pm$0.33 & 0.48$\pm$0.03 & 51$\pm$9 & 11.8$\pm$0.2 & [11]\\
        \hline
    \end{tabular}
\end{table*}

\section{Conclusions}

The {\sl Gaia} TGAS catalogue and the catalogue of Astraatmadja \&
Bailer-Jones (2016) provide parallaxes and distances to 16 CVs and
related objects. The Gaia parallax to SS\,Cyg is consistent with that
determined using a VLBI parallax measurement and is in spectacular
agreement with the distance implied by the DIM when modeling the
observed outburst behaviour of SS\,Cyg. The TGAS distances to CVs have
allowed us tentatively to explore the relationship between absolute
visual magnitude, accretion disc state and orbital period and how the
sub-types of CVs relate. We have also shown that accurate X-ray
luminosities allow the accretion state of the system to be determined,
whether that be a high or low state accretion disc or a magnetic
propeller, as well as determining the mass accretion rate in
quiescence and at the transition to an optically-thick boundary layer
in outburst. We have demonstrated the value of accurate parallaxes
combined with ultraviolet observations in determining the mass of CV
white dwarfs and the secular mass transfer rates, both key parameters
in understanding the evolution of CVs. We have also estimated the
critical mass accretion rate for the NL variables and the DN in our
sample, and compared these values with the observational estimates of
the mass accretion rates based on the {\sl Gaia} distances and
historic AAVSO data. We find good general agreement although it is
somewhat marginal for a few systems. The second {\sl Gaia} data
release will provide parallaxes to many hundreds of CVs, and thus
provide the data base necessary to extend these fundamental studies to
CVs spanning the entire range of orbital period, mass transfer rate,
and sub-type.

\section{Acknowledgements}

This work has made use of data from the ESA space mission {\sl Gaia},
processed by the {\sl Gaia} Data Processing and Analysis Consortium
(DPAC) and we thank everyone who has been involved with the
mission. We thank all the observers of the AAVSO for their tireless
and important work. Armagh Observatory and Planetarium is core funded
by the Northern Ireland Executive through the Dept for Communities. PW
is supported by a STFC consolidated grant (ST/L000733/1). The research
leading to these results has received funding from the European
Research Council under the European Union's Seventh Framework
Programme (FP/2007-2013) / ERC Grant Agreement n. 320964 (WDTracer). MRS thanks for support from Fondecyt (grant 1141269). We thank the referee for a careful reading of the manuscript.

\appendix

\section{Background notes on individual sources}

{\bf V603 Aql:} With a peak brightness of $V\sim$0.5, Nova Aql 1918
(V603 Aql), is the brightest Galactic nova to ever have been
identified (Strope, Schaefer \& Henden 2010). It took 20 years for
V603 Aql to fade to its pre-outburst brightness, but has continued to
decline in brightness (Johnson et al. 2014). The astrometric distance
determined using the HST fine guidance sensor is 249$^{+9}_{-8}$ pc
(Harrison et al. 2013) which compares with 328$^{+60}_{-29}$ pc
derived from nebular expansion studies (Downes \& Duerbeck 2000). It
has an orbital period of 3.3 hr (Kraft 1964) and a low binary
inclination ($i\sim20^{\circ}$, Patterson et al. 1993). Observations
made using {\sl ASCA}, {\sl RXTE} and {\sl Chandra} show flux
variability which is likely caused by the changing maximum temperature
of the plasma (Mukai \& Orio 2005). A distance of 110$\pm$6 pc is
determined using linear polarisation (Barrett 1996). The distance
obtained using {\sl Gaia} is 328$\pm$78 pc.\medskip

{\bf RR Pic:} RR Pic was seen as a nova in 1925 reaching a peak of
$V\sim$1.0 and showed substantial variations in its brightness on the
decline to quiescence of V$\sim$12.2 (Strope et al. 2010). An orbital
period of 3.48 h was identified by Vogt (1975) and the binary
inclination is 60$^{\circ}<i<80^{\circ}$ (Riberio \& Diaz (2006). RR
Pic was observed using {\sl ROSAT} and found to be a moderately bright
X-ray source (van Teeseling et al. 1996) whilst observations made
using {\sl Chandra} show evidence for various emission lines in the
X-ray spectrum (Pekon \& Balman 2008). The astrometric distance
determined using the HST fine guidance sensor is 521$^{+54}_{-45}$ pc
(Harrison et al. 2013). Using models of the expansion of the
surrounding nebula, Gill \& O'Brien (1998) estimate a distance to RR
Pic of 600$\pm$60 pc. A distance of 240$\pm$65 pc is determined using
linear polarisation (Barrett 1996). The distance obtained using {\sl
  Gaia} is 388$\pm$88 pc.\medskip

{\bf HR Del:} HR Del was discovered as a nova in 1967 and had a peak
brightness of $V\sim$3.6 and returned to its quiescence mag
$V\sim$12.1 after about a decade (Strope et al. 2010). K\"{u}rster \&
Barwig (1988) determined an orbital period of 5.14 d and a binary
inclination of $i\sim40^{\circ}$. Harman \& O'Brien (2003) derived a
distance of 970$\pm$70 pc using models of the expansion of the nova
shells.  Although a strong UV source, HR Del was only marginally
detected in hard X-rays using {\sl EINSTEIN} (Hutchings 1980).  The
distance obtained using {\sl Gaia} is 560$\pm$170 pc.
\medskip 

{\bf TT\,Ari:} TT\,Ari was one of the earliest CVs to be discovered
and have its orbital period to be determined (3.3 hr, Cowley et
al. 1975). It was detected as a strong hard X-ray source using {\sl
  Einstein} (Cordova, Mason \& Nelson 1981). Further spectroscopic and
photometric observations indicated it undergoes rapid fading events in
the optical which are typical of VY Scl type CVs (Shafter et al
1985). Shafter et al. give a lower limit of 200 pc to its distance and
find a relatively low binary inclination ($i\sim30^{\circ}$).
G\"ansicke et al. (1999) analysed {\sl IUE} far-ultraviolet and
optical spectroscopy obtained in 1982/83 during a deep low state, and
detected the photospheric signatures of both the white dwarf and the
M-dwarf companion. Based on the derived spectral type of the companion
and the optical flux, they estimated a distance of $d=335\pm50$\,pc. A
distance of 209$\pm$64 pc is determined using linear polarisation
(Barrett 1996). The distance obtained using {\sl Gaia} is 228$\pm$30
pc.\medskip

{\bf IX Vel:} Spectroscopic observations by Wargau et al. (1983)
showed IX Vel to have spectra typical of a NL CV and giving a
preliminary orbital period of 4.5 hrs. By fitting the $K$ band light
curve Linnell et al. (2007) determined a binary inclination of
57$\pm2^{\circ}$.  No outbursts were detected over several years
(Garrison et al. 1984) confirming IX Vel as a NL CV above the orbital
period gap. Duerbeck (1999) gives a distance determined using
Hipparchos of 96$^{+10}_{-8}$ pc. X-ray observations of IX Vel made
using {\sl ROSAT} at two epochs show it as a strong X-ray source
whilst further X-ray observations were made using {\sl ASCA} (Baskill,
Wheatley \& Osborne 2005). A distance of 81$\pm$44 pc is determined
using linear polarisation (Barrett 1996). The distance obtained using
{\sl Gaia} is 89$\pm$3 pc.\medskip

{\bf UX UMa:} UX UMa is a bright ($V\sim$13) eclipsing CV with an
orbital period of 4.7 h (Walker \& Herbig 1954) and is the archetypal
member of the NL group of systems which are always in a high accretion
state (Walker \& Herbig 1954).  Baptista et al. (1995) determine a
distance of 345$\pm$34 pc using photometric colours and estimates of
the size of the accretion disc (Baptista et al. 1995) and derive a
binary inclination of $i=71.0^{\circ}\pm0.6^{\circ}$.  Observations
made using {\sl XMM-Newton} show that whilst there is no eclipse in
soft X-rays there is an eclipse in hard X-rays indicating the latter
are emitted in the boundary layer between the accretion disk and the
white dwarf (Pratt et al. 2004). A distance of 333$\pm$11 pc is
determined using linear polarisation (Barrett 1996). The distance
obtained using {\sl Gaia} is 264$\pm$30 pc.\medskip

{\bf V3885 Sgr:} Cowley et al. (1977) identified V3885 Sgr as a bright
($V\sim$10.3) CV which did not show outbursts. Hartley et al. (2005)
derived an orbital period of 5.0 h and detected spiral waves in the
acretion disc, whilst its distance determined via Hipparchos data is
110$^{+30}_{-20}$ pc (Duerbeck 1999). Linnell et al. (2009) determine
a binary inclination of $i\sim65^{\circ}$ by modelling optical and UV
spectra.  van Teeseling \& Verbunt (1994) showed V3885 Sgr to be
moderately bright in {\sl ROSAT} data and found that it showed a
stable flux during these observations and in comparison with
observations made using {\sl EINSTEIN} and {\sl EXOSAT}.The distance
obtained using {\sl Gaia} is 136$\pm$8 pc.\medskip

{\bf RX And:} RX And was discovered through regular optical outbursts
and was found to be a binary with a period of 5.1 h by Kraft
(1962). Observations made by members of the AAVSO show outbursts every
few weeks, but also epochs where it is bright for several months (and
typical of Z Cam stars) and epochs where it is faint for several
months (and typical of VY Scl stars, Schreiber, G\"{a}nsicke, Mattei
2002). Shafter (1983) determines a binary inclination of
$i\sim51\pm9^{\circ}$ and Sion et al. (2001) determine a distance of
200 pc based on {\sl HST} spectra and assumptions on the mass of the
white dwarf.  RX And was detected in X-rays using the {\sl EINSTEIN}
satellite (Eracleous, Halpern, Patterson 1991) and as a weak source
using {\sl ROSAT} (Verbunt et al. (1997). A distance of 318$\pm$55 pc
is determined using linear polarisation (Barrett 1996). The distance
obtained using {\sl Gaia} is 185$\pm$24 pc.\medskip

{\bf HL CMa:} HL CMa showed outbursts approximately every 15 days
(Chlebowski, Halpern \& Steiner 1981) although it can also show
standstill events (V$\sim$12) (Kato 2002). As such it appears to be a
Z Cam type CV.  Phase resolved spectroscopy made by Hutchings et
al. (1981) showed an orbital period $\sim$5 hrs and a binary
inclination of $i\sim45^{\circ}$, with the orbital period being
refined to 5.2 hrs (Still et al. 1999).  A distance of 220$\pm$10 pc
is determined using linear polarisation (Barrett 1996). The distance
obtained using {\sl Gaia} is 310$\pm$43 pc.\medskip

{\bf AH Her:} AH Her was identified as a U Gem type-variable through
its outbursts (Petit 1960) and has an orbital period of 5.9 h (Moffat
\& Shara 1984). The secondary star has an early to mid K spectral
type, and has a binary inclination of $i\sim46\pm3^{\circ}$ (Horne,
Wade \& Szkody 1986). It has shown prolonged periods of standstill
indicating it is Z Cam type CV (Simonsen 2011).  Thorensten (2003)
determined a distance of 660$^{+270}_{-200}$ pc via parallax
determinations. AH Her was detected in the {\rosat} all-sky survey at
a low rate (Verbunt et al. 1997). Simultaneous optical and UV ({\it
  IUE}) observations show that the UV flux follows the optical during
an outburst (Verbunt et al. 1984). A distance of 556$\pm$110 pc is
determined using linear polarisation (Barrett 1996). The distance
obtained using {\sl Gaia} is 325$\pm$47 pc.\medskip

{\bf SS Cyg:} SS Cyg is one of the best studied CVs in the sky being
discovered in 1896 (Wells 1896) having an orbital period of 6.6 h
(Walker \& Chincarini 1968).  It shows outbursts approximately every
50 days reaching a peak magnitude of $V\sim$8 (Cannizzo \& Mattei
1992). Friend et al. (1990) found a binary inclination of
$i=37^{\circ}-53^{\circ}$ assuming a white dwarf mass 0.55--1.2 \Msun,
while Bitner et al. (2007) find $i=45^{\circ}-56^{\circ}$ and a white
dwarf mass 0.81$\pm$0.19 \Msun.  The distance to SS Cyg has been the
subject of much debate. Using the {\sl HST} fine guidance sensor to
measure its parallax, Harrison et al. (2004) revised their previous
distance to SS Cyg from 166$\pm$12 pc to 152$\pm$9 pc. Both distances
were at odds with a parallax determined using the VLBI of 114$\pm$2 pc
(Miller-Jones et al. 2013). Although Nelan \& Bond (2013) re-analysed
the {\sl HST} data and found a distance of 120.5$\pm$5.7 pc, this
analysis was questioned by Harrison \& McArthur (2016) who now
conclude the distance to SS Cyg is 137$\pm$4 pc (which remains at odds
with the VLBI distance). Wheatley, Mauche \& Mattei (2003) shows that
although there is a sudden burst of hard X-rays $\sim$1 d after the
start of the optical outburst, they are suppressed shortly afterwards,
only increasing again as the optical outburst approaches quiescence. A
distance of 67$\pm$17 pc is determined using linear polarisation
(Barrett 1996). The distance obtained using {\sl Gaia} is 117$\pm$6
pc.\medskip

{\bf Z Cam:} Z Cam is the prototype of CVs which normally show regular
outbursts (similar to SS Cyg) but which show `standstills' which can
last for many months (e.g. Mayall 1965). Z Cam stars have orbital
periods longer than 3 hr (Z Cam has an orbital period of 6.96 h,
Kraft, Krzem\'{i}nski \& Mumford 1969), placing them above the upper
limit of the 2--3 hr period gap (Simonsen et al. 2014). Shafter (1983)
determines a binary inclination of 57$^{\circ}\pm11^{\circ}$. The
discovery of a shell of material around the binary implies a nova
outburst in the distant past (Shara et al. 2007). Thorstensen (2003)
determine a parallax of 163$^{_+68}_{-38}$ pc. Baskill, Wheatley \&
Osborne (2001) found using {\sl ASCA} data that the observed flux was
an order of magnitude higher at the point where it transitioned
between quiescence and an outburst. The distance obtained using {\sl
  Gaia} is 219$\pm$20 pc.\medskip

{\bf BV Cen:} BV Cen was found to show one magnitude variations on
plates taken in 1928 (Waterfield 1929) and later classed as a U Gem
type CV (Kraft \& Luyten 1965). However, it was two decades later when
the orbital period was shown to be 14.6 h which made it one of the
longest period CVs (Vogt \& Breysacher 1980). However, Menzies,
O'Donoghue \& Warner (1986) show that BV Cen has a brighter absolute
magnitude than expected for a regular DN and suggest it may have
undergone a nova eruption. Watson et al. (2007) determine a binary
orbital period of $i=53^{\circ}\pm4^{\circ}$. BV Cen was recently
observed using {\sl Suzaku} (Xu et al. 2016) and has a distance of 435
pc determined by modelling {\sl FUSE} spectra (Sion et al. 2007). The
distance obtained using {\sl Gaia} is 344$\pm$65 pc.
\medskip

{\bf AE Aqr:} AE Aqr has long been known as a bright variable, with
Gaposchkin (1949) noting it was a member of the U Gem class. However,
it was many years before its true nature was revealed, first when
Payne-Gaposchkin (1969) identified the binary orbital period at 9.88
hrs and then when Patterson (1979) found evidence for a stable optical
period of 33.08 sec. The shorter period is due to the spin period of
the accreting white dwarf and is classed as an intermediate polar,
whose field strength is sufficiently high ($B\sim1-10$MG) to dominate
the dynamics of the accretion flow at some distance from the white
dwarf photosphere. The unique characteristics of this system led to an
interpretation in which most of the mass transfer flow is propelled
away from the white dwarf by its rapidly rotating magnetosphere (Wynn,
King \& Horne 1997). Hill et al. (2014) determined a binary
inclination of $i\sim50^{\circ}$. A distance of 102$^{+42}_{-23}$ pc
was determined using Hipparchos data (Friedjung 1997). Simultaneous
observations were made of AE Aqr using NuStar and {\swift} in 2012
(Kitaguchi et al. 2014). A distance of 155$\pm$55 pc is determined
using linear polarisation (Barrett 1996). The distance obtained using
{\sl Gaia} is 91$\pm$3 pc.

{\bf QU Car:} Spectroscopic observations showed that QU Car is a long
period (10.9 h) binary with a hot accretion disc and a binary
inclination $i<60^{\circ}$. (Gilliland \& Philips 1982). The distance
to QU Car is not well determined but Gilliland \& Philips (1982)
measured a distance $>$500 pc, whilst Drew et al. (2003) indicated it
could lie at 2 kpc, which would make it a very luminous CV with a
possible Carbon star secondary.  QU Car was the target of two pointed
observations by {\xmm}. The distance obtained using {\sl Gaia} is
410$\pm$86 pc.\medskip

{\bf V Sge:} V Sge is an eclipsing binary with a period of 12.3 hr
which shows epochs of a quasi-cycling behaviour when it can change its
optical brightness by $\sim$3 mag on a timescale of weeks (Herbig et
al. 1965). Its nature has been long debated but appears to consist of
two hot stars, with the hotter star very close to filling its Roche
lobe with a binary inclination of $i\sim72^{\circ}$ (Lockley et
al. 1999). V Sge has a parallax distance of 320$_{-260}^{+\infty}$ pc
(van Altena, Lee \& Hoffleit 1995). {\sl ROSAT} observations showed
that V Sge during a low optical states appears as a super-soft X-ray
source (Greiner \& van Teeseling 1998). A distance of 56$\pm$56 pc is
determined using linear polarisation (Barrett 1996). The distance
obtained using {\sl Gaia} is 760$\pm$210 pc.

\end{document}